\documentclass[traditabstract]{aa}

\usepackage{epsfig,graphicx,natbib}
\usepackage{longtable}
\usepackage{amssymb}
\usepackage{amsmath}

\def\nH2{\hbox{$n_\mathrm{H_2}$}}
\def\kms{\hbox{km\,s$^{-1}$}}
\def\PKS1830{\hbox{PKS\,1830$-$211}}
\def\cm-2{\hbox{cm$^{-2}$}}
\def\cm3{\hbox{cm$^{-3}$}}
\def\fH2{\hbox{$f_{\rm H2}$}}

\def\H0{\hbox{$H_0$}}

\begin{document}

\title{All good things come in threes: the third image of the lensed quasar \PKS1830}

\author{S.~Muller \inst{1}
\and S.~Jaswanth \inst{1}
\and C.~Horellou \inst{1}
\and I. Mart\'i-Vidal \inst{2,3}
}
\institute{Department of Space, Earth and Environment, Chalmers University of Technology, Onsala Space Observatory, SE-43992 Onsala, Sweden
\and Observatori Astron\`omic, Universitat de Val\`encia, Parc Cient\'ific, C. Catedr\`atico Jos\'e Beltr\'an 2, 46980 Paterna, Val\`encia, Spain
\and Departament d'Astronomia i Astrof\'isica, Universitat de Val\`encia, C. Dr. Moliner 50, 46100 Burjassot ,Val\`encia, Spain
}

\date {Received / Accepted}

\titlerunning{The third image of \PKS1830}
\authorrunning{S. Muller et al.}

\abstract{Strong gravitational lensing distorts our view of sources at cosmological distances but brings invaluable constraints on the mass content of foreground objects and on the geometry and properties of the Universe. We report the detection of a third continuum source toward the strongly lensed quasar \PKS1830 in ALMA multi-frequency observations of high dynamic range and high angular resolution. This third source is point-like and located slightly to the north of the diagonal joining the two main lensed images, A and B, $0.3\arcsec$ away from image B. It has a flux density that is $\sim 140$ times weaker than images A and B and a similar spectral index, compatible with synchrotron emission. We conclude that this source is most likely the expected highly de-magnified third lensed image of the quasar. In addition, we detect, for the first time at millimeter wavelengths, weak and asymmetrical extensions departing from images A and B that correspond to the brightest regions of the Einstein ring seen at centimeter wavelengths. Their spectral index is steeper than that of compact images A, B, and C, which suggests that they arise from a different component of the quasar. Using the GravLens code, we explore the implications of our findings on the lensing model and propose a simple model that accurately reproduces our ALMA data and previous VLA observations. With a more precise and accurate measurement of the time delay between images A and B, the system \PKS1830\ could help to constrain the Hubble constant to a precision of a few percent.}

\maketitle

\section{Introduction}

The theory of general relativity describes gravity as a geometric property of the space-time continuum. The presence of a mass changes the local curvature of the Universe and can bend the trajectory of photons, possibly leading to distorted images of background sources. Over cosmological distances, even a small deflection of the photons can result in a different path-length and an apparent time delay between lensed images, related to the Hubble constant, \H0. Gravitational lensing therefore offers a unique opportunity to assess mass distributions in the Universe and measure \H0 via the so-called time-delay cosmography method (\citealt{ref64} and, e.g., \citealt{won20} and results from the H0LiCOW collaboration).

The relative positions of the source, the lens, and the observer set the number of lensed images, their magnifications, and the time delays between them (e.g., \citealt{sch92}). In the special geometrical case where the source, lens, and observer are closely aligned, the resulting image of the source is a perfect Einstein ring. More generally, a gravitational lens always produces an odd number of images, partially or well separated. One of them is always located close to the center of the lens and, in contrast to other images, is highly de-magnified (by a factor $\sim$10$^2$ to $10^4$). This central image brings important information: its position and de-magnification ratio can put strong constraints on the lens geometry (lens position, ellipticity of the dark matter halo) and on the lens central mass distribution (see e.g., \citealt{kee03,tam15, won15}).

Detections of the central image are rare and difficult (e.g., \citealt{win04,qui16}). They require high sensitivity in order to overcome the strong demagnification, high angular resolution in order to resolve the different components (typically separated by a few arcseconds in the sky), and high dynamic range ($>100-1000$). In addition, confusion can arise between the third image, that is, emission from the background source, and photons emitted by the lens itself. Finally, propagation effects (e.g., free--free absorption through the lens, or extinction and reddening due to dust) can also affect the observations.

The system \PKS1830\ is a remarkable case of strong gravitational lensing. It consists of a radio-loud quasar at $z=2.5$ (\citealt{lid99}), a foreground lens in the form of a nearly face-on $z=0.89$ spiral galaxy (\citealt{wik96,win02,koo05}) which is also responsible for remarkable molecular absorption (e.g., \citealt{wik96, mul11, mul14}), and a second intervening galaxy at $z=0.19$ with H\,I and OH absorption (\citealt{lov96,all17}). However, this second galaxy is expected to have a negligible effect on the lensing (\citealt{win02,sri13}). The resulting image of the quasar appears at centimeter wavelengths as two highly magnified compact components (hereafter labeled images A and B, to the northeast and southwest, respectively) separated by $\sim 1\arcsec$ and embedded in a full Einstein ring (\citealt{sub90,jau91,nai93}). Another component close to image B was tentatively identified as the third image by \cite{nai93}.

Although many observations have been performed toward \PKS1830\ at different wavelengths (radio, optical, X- and $\gamma$-rays), there has been so far no convergence toward a single lens model. One of the major sources of debate and uncertainty is the lens position (Table~\ref{tab:astrometry-literature}, see also, e.g., \citealt{sri13}). \PKS1830\ is located behind the Galactic plane, close to the direction of the Galactic Center ($b$=$-5.7^{\circ}$), meaning that optical observations suffer from high extinction by dust and contamination by Galactic stars, making a robust optical identification difficult (e.g., \citealt{win02,cou02}). In radio--cm observations, the third image is tentatively identified (component E by \citealt{sub90}), with a flux density more than one hundred times lower than that of image A (\citealt{nai93}). However, the variability of \PKS1830\ prevents the use of data obtained at different epochs, making it difficult to measure spectral indices to confirm the identification. In addition, there are some ambiguities in the radio--cm images with possible jet-knot extensions from image B. Similarly, very-long-baseline interferometric (VLBI) observations show strong intrinsic variability of the quasar images (\citealt{gar97, gui99}), making it difficult to disentangle the jet--core sub-components. The VLBI observations are also limited in terms of dynamic range, making it difficult to detect the faint third image.

Concerning measurements of the time delay between images A and B, several methods have been deployed with various success, from radio monitoring (\citealt{vanomm95,lov98}) to analysis of molecular absorption variations (\citealt{wik01}) and $\gamma$-ray time-series (\citealt{bar11,abd15,bar15}), with results pointing toward a most likely value of between 20 and 30 days.
Finally, even the differential magnification ratio ($\mu_{\rm A}/\mu_{\rm B}$) is uncertain because the instantaneous flux density ratio A/B is varying with time due to the intrinsic variability of the quasar modulated by the time delay. Values in the range 1-2 have been observed (e.g., \citealt{vanomm95,mul08, mar13, mar20}). In principle, micro-lensing on one line of sight could also affect the ratio for some time (e.g., on a 1 year timescale).

Here, we report new millimeter observations with the Atacama Large Millimeter/submillimeter Array (ALMA) that allowed us to unambiguously identify the third image of \PKS1830\ and refine the lens model of the system.

\section{Observations} \label{sec:obs}

Our ALMA observations of \PKS1830\ were carried out on 2019 July 10 and 11 in bands 4 ($\sim 150$~GHz) and 6 ($\sim 230$~GHz), and on 2019 July 28 in bands 5 ($\sim 180$~GHz) and 7 ($\sim 280$~GHz). The array was in an extended configuration, with longest baselines up to 8--14~km depending on the date. This resulted in a synthesized beam of between 60~mas and 20~mas. The largest recoverable angular scale (i.e., due to the filtering of extended emission by the interferometer) is larger than the extent of the Einstein ring, and therefore we do not expect to miss extended emission. A summary of the observations is given in Table~\ref{tab:obsdata}.

The correlator was configured to cover four independent spectral windows, each of 1.875~GHz in width, with a spectral resolution $\sim 0.5$--1~MHz adapted for the study of the molecular absorption lines arising from the $z=0.89$ intervening lens galaxy (e.g., \citealt{mul14}). All the absorption features were flagged to image the continuum emission.

The data calibration was done within the CASA\footnote{http://casa.nrao.edu/} package, following a standard procedure. The bandpass response of the antennas was calibrated from observations of the bright quasar J\,1924$-$292 (3.8 to 2.5~Jy with increasing frequency), which was also used for the absolute flux calibration. Conservatively, the flux accuracy is better than 5\% at bands 4, 5, and 6, and better than 10\% at band 7, as retrieved from the ALMA observatory flux monitoring data of J\,1924$-$292 at the date of observations.
The quasar J\,1832$-$2039, $\sim 0.3$~Jy and about half a degree away from \PKS1830\ in the sky, was observed every 5--7~min for gain (phase and amplitude) calibration.

After the standard calibration, the visibility phases were furthermore self-calibrated on the continuum of \PKS1830, applying phase-only gain corrections at each 6\,s integration. This self-calibration step  dramatically improved the quality of the data by reducing the atmospheric decoherence. In particular, the dynamic range (i.e., the ratio of peak flux to rms noise level in the image) improved by a factor of two to three after self-calibration, and reached values $\gtrsim 1000$, which is about the high-end expectation for ALMA observations \footnote{see, e.g., the ALMA Technical Handbook for Cycle 7, \S\,10.5.1.}. This is possible due to the strong flux density of the A and B images of  \PKS1830, which allows us to derive self-calibration gain corrections on the shortest integration interval, although the dynamic range is still limited, for example, by antenna position uncertainties, potential residual calibration issues, and digit quantization in the correlator. 

The final images of the continuum emission (Fig.\,\ref{fig:ContImages}) were produced for each band separately, from a H\"ogbom-Clean multi-frequency synthesis deconvolution of the interferometric visibilities, with a pixel size of 5~mas.

\section{Results} \label{sec:results}

\begin{figure*}[t] \begin{center}
\includegraphics[width=\textwidth]{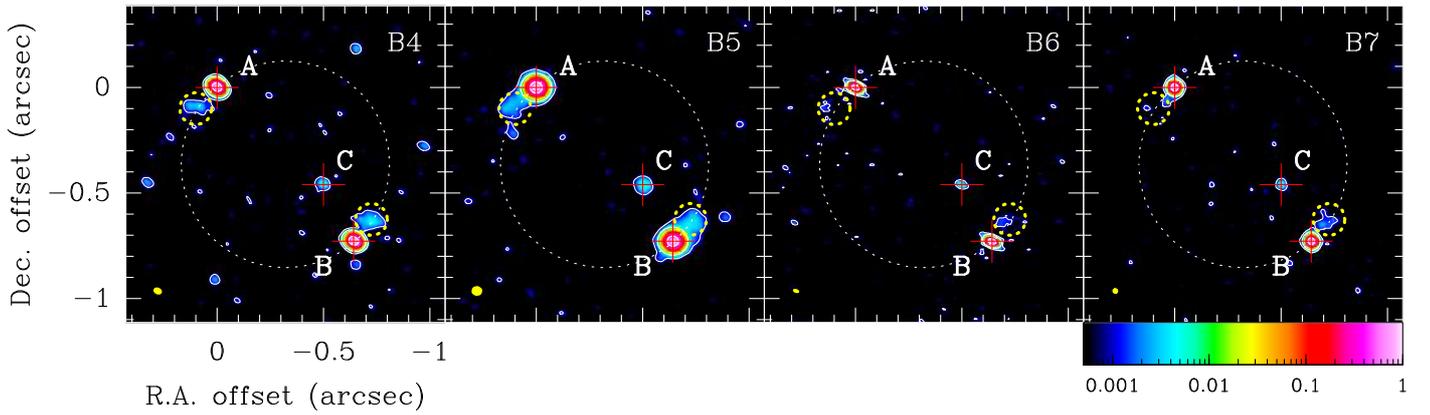}
\caption{Continuum images of \PKS1830\ obtained with ALMA in bands 4, 5, 6, and 7. Intensities are normalized to the peak of each image. The color wedge shown at the bottom of the extreme right box is the same for all images. Contour levels are drawn at 0.1\%, 1\%, and 50\%. The positions of images A, B, and C are marked with red crosses. The weak extensions departing from images A and B are indicated by the yellow dashed circles. The center of the white dashed circle corresponds to the halfway point between images A and B. The synthesized beam is shown as the yellow ellipse in the bottom-left corner of each box.}
\label{fig:ContImages}
\end{center} \end{figure*}

The maps of \PKS1830\ in the four observed ALMA bands (Fig.\,\ref{fig:ContImages}) reveal {\em (i)} the two images A and B, as two bright point-like sources, {\em (ii)} weak asymmetrical extensions departing from these images, to the southeast for image A and to the northwest for image B, and {\em (iii)} an additional faint and isolated third point-like source, hereafter denoted image C, located slightly north above the line joining images A and B, about 0.3$\arcsec$ away from image B. The positions and flux properties of the three images A, B, and C are listed in Table~\ref{tab:info-images}.

The location of the third point-like source, its flux ratio relative to images A and B, and its similar spectral index point toward its identification as the expected third, de-magnified lensed image of \PKS1830. We note that its position is consistent with ``feature E'' identified on the VLA maps by \cite{sub90} and \cite{nai93}, and with the deconvolved position of the lens galaxy by \cite{mey05} (see Table~\ref{tab:astrometry-literature}). The ALMA identification of the third image finally removes any confusion around the nature of feature E and the position of the lens galaxy (\citealt{cou02,win02,mey05}), hence allowing a major improvement of the lens model.

The weak extensions departing from images A and B are the counterparts of the brightest regions of the Einstein ring seen at centimeter wavelengths (\citealt{sub90}). These are more conspicuous in ALMA bands 4 and 5 than in bands 6 and 7, and we estimate spectral indices $\alpha = -1.28 \pm 0.15$ and $-1.18 \pm 0.15$ toward A and B extensions, respectively. These spectral indices are significantly steeper than those of the point-like images A, B, and C, which indicates that the extensions correspond to a different source component. The extension departing from image B is about 1.4 times brighter than the one departing from image A. However, we note that the flux ratio of the extensions may vary with time in a way that is not necessarily related to the variations of the flux ratio of the images, A/B.

\begin{figure}[h] \begin{center}
\includegraphics[width=8.8cm]{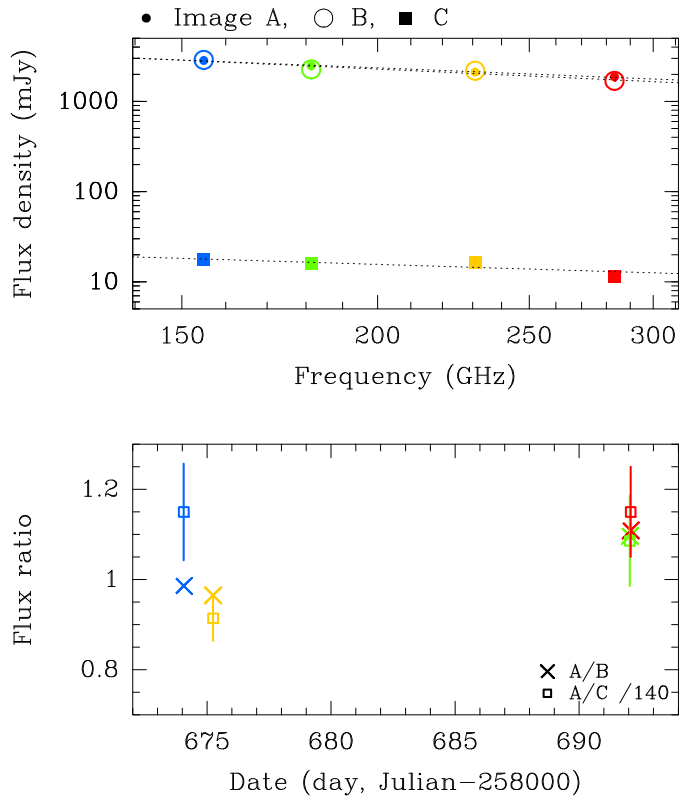}
\caption{Flux density of the three images of \PKS1830\ as a function of frequency and power-law fit of the spectral indices (top panel), and flux density ratios as a function of time (bottom panel). The color code is the same for both figures.}
\label{fig:SpectralIndex}
\end{center} \end{figure}

\begin{table*}[ht]
\caption{Properties of the three lensed images of \PKS1830\ from the ALMA 2019 data.}
\label{tab:info-images}
\begin{center} \begin{tabular}{cccccc}
\hline
Image & R.A. offset & Dec. offset & Flux density at 300~GHz & Flux normalized & Spectral index \\
      & (mas) & (mas) & (mJy) & to image A & $\alpha$ \\
\hline
A & 0.0 & 0.0 & $1776 \pm 36$ & 1.0 & $-0.698 \pm 0.045$ \\
B & $-641.946 \pm 0.042$ & $-728.082 \pm 0.033$ & $1658 \pm 165$ & $0.93 \pm 0.09$ & $-0.79 \pm 0.21$ \\
C & $-500.1 \pm 2.1$ & $-459.6 \pm 1.6$ &$12.6 \pm 1.6$ & $0.0071 \pm 0.0009$ & $-0.53 \pm 0.28$ \\ 
\hline
\end{tabular}\end{center}
\tablefoot{Flux densities come from the global fit of the spectral index, see Fig.\,\ref{fig:SpectralIndex}. The uncertainties do not take into account $\sim 10$\% uncertainty on the absolute flux calibration, but carry intrinsic source variability, with time variations of 15\% observed in the flux density ratio A/B between 10 and 28 of July 2019 (Table~\ref{tab:FluxRatios}). The spectral index is defined as $F = F_0 \times (\nu / \nu_0)^{\alpha}$.}
\end{table*}

\section{Revising the lens model for \PKS1830 }

Armed with the new constraint of the unambiguous position of the third image, we can revise the lens model for \PKS1830. For this, we used the GravLens code\footnote{http://www.physics.rutgers.edu/$\sim$keeton/gravlens/} (\citealt{kee01}) in successive steps, as described below. 

First, we modeled the source and images as point-like objects, that is, ignoring the weak extensions. A softened isothermal ellipsoidal model was adopted for the lens mass distribution (e.g., \citealt{kor94} and see App.\,\ref{app:lens}). In addition to the redshifts of the source and lens, the lens model has therefore eight free parameters: the position of the (point-like) source and the position, scale mass, softening length, ellipticity, and position angle of the softened isothermal ellipsoidal lens. We took the position of the third image as an initial guess for the center of the lens, because both are expected to be very close. More details are given in Appendix\,\ref{app:lensmodel}. The best-fit model was obtained by minimizing the $\chi$-square in the source and image plane, successively (see \citealt{kee01}). The values of input and output parameters are listed in Table~\ref{tab:lensing-model_pointlike}. The outcome of the fit is shown in the form of Monte-Carlo degeneracy parameter plots ("corner plots") in Fig.\,\ref{fig:mcmc}. 

Our goal was not to achieve the ultimate modeling of the system, but to build one plausible model that could satisfy the existing VLA and ALMA observations, and in particular the new constraint imposed by the identification of the third image. As previously mentioned, we considered that we could neglect line-of-sight structures, starting with the second absorber at $z=0.19$ (see Appendix\,\ref{app:2ndabsorber}). 

We also refrained from trying to constrain the value of \H0, because there is currently no precise and accurate determination of the time delay $t_{\rm AB}$ . Instead, we ran the modeling with two values of \H0, the first one derived from {\em Planck} data (\citealt{Planck18}) and the second from modeling strong gravitational lens systems (\citealt{won20}). Those two values are close enough (although statistically in tension with each other) so that the corresponding time delays between the images do not differ by more than $\sim 10$\%. Reciprocally, the exercise shows that a precise and accurate measurement of the time delays toward \PKS1830\ could help constrain \H0\ possibly to a precision of a few percent, although sources of systematic errors (such as line-of-sight structures, shear degeneracy, lens model choice, see, e.g., \citealt{won20,mil20}) would have to be carefully explored, which is beyond the scope of this paper.

\begin{figure*}[h] \begin{center}
\includegraphics[height=5.5cm]{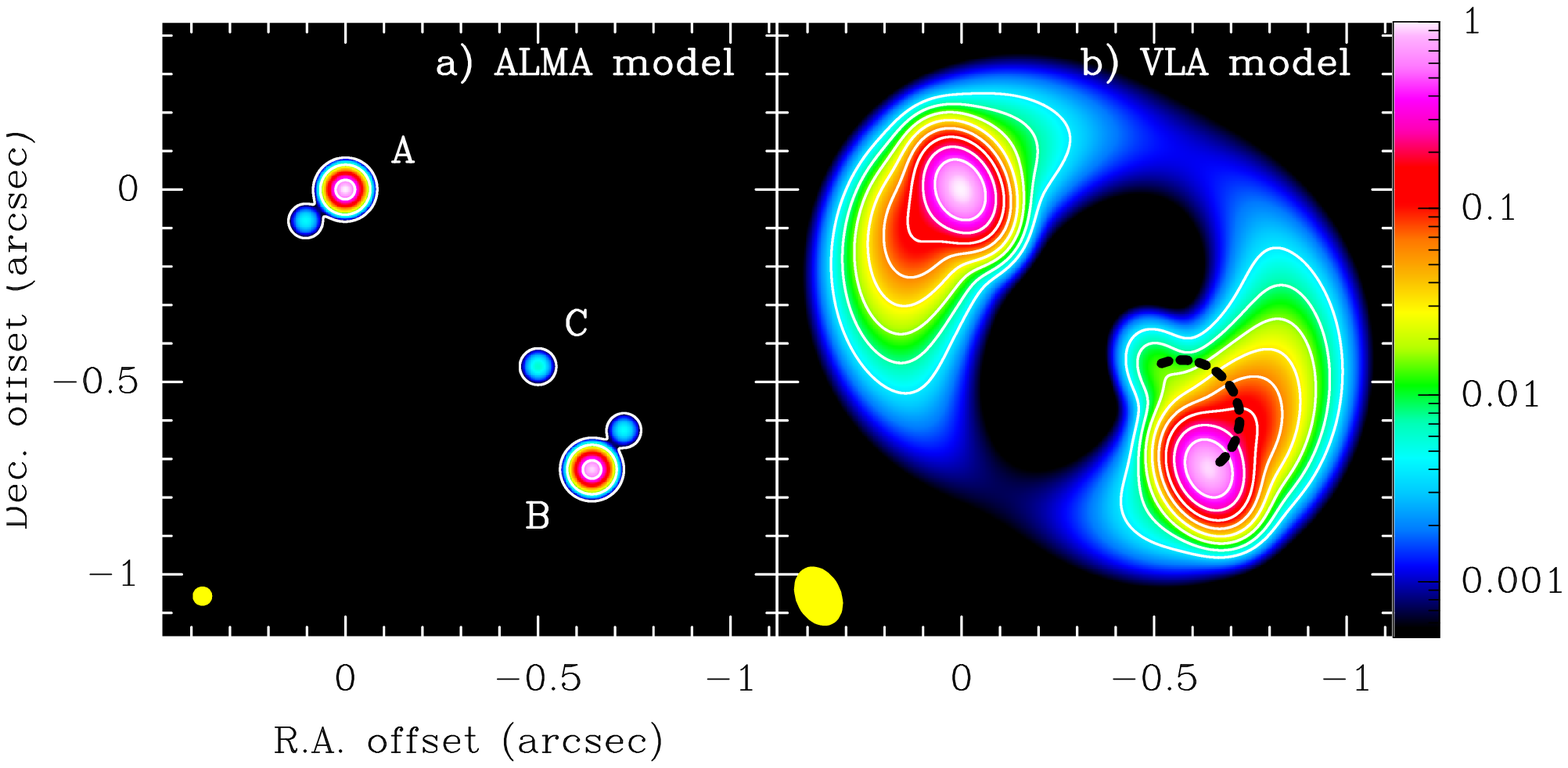}
\includegraphics[height=5.5cm]{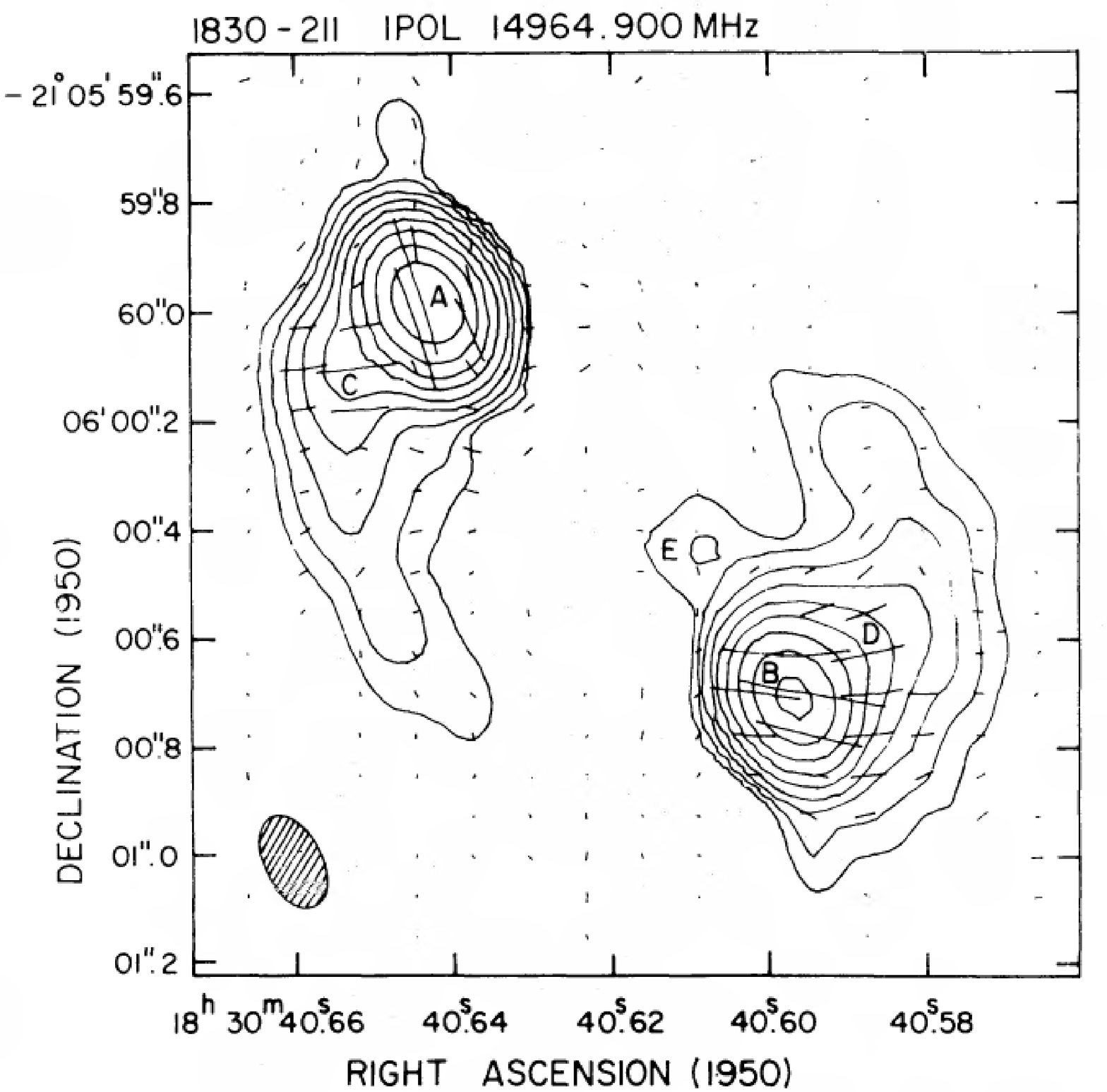}
\caption{Results from the ALMA and VLA lens models (see Table~\ref{tab:ext_sou_models}). Images are normalized to the peak intensity and smoothed to the beam (shown in bottom-left corner) given hereafter. (a) ALMA model with two point-like sources, smoothed to a resolution of 30~mas. Contour levels at 0.5, 0.01, and 0.0005. (b) VLA model, with one point-like and one extended source, smoothed to the same beam as the VLA-15\,GHz observations by \cite{sub90}: $161 \times 117$~mas$^2$, with a position angle of 26$^{\circ}$. Contour levels every 1/2$^n$, with $n=1$ to 10. In our simulations, a thin jet would follow an arclet feature between images B and C, as indicated by the dashed curve. Right: VLA-15\,GHz observations by \cite{sub90}. We note that their labels of the different features are not the same as ours: in this image, labels C and D mark the extensions from images A and B, and label E corresponds to the third image.}
\label{fig:finalmodels}
\end{center} \end{figure*}

For the lens model to reproduce the weak extensions and, subsequently the Einstein ring seen at centimeter wavelengths, we have to add components to the source. Although the lensing is achromatic, these components (such as a core and a jet) may have different spectral indices, and therefore we may need to adopt different source models to reproduce the ALMA and VLA observations. First, keeping the same lens model as derived above, we considered two point-like images at the positions of the weak extensions seen by ALMA (center of the yellow dashed circles in Fig.\,\ref{fig:ContImages}), and searched for their corresponding point-like source in the source plane. We found that this second point-like source would fall at a position $(-333.33,-341.55)$~mas relative to image A, that is, a separation of 55~mas ($\sim 0.5$~kpc at the redshift of the quasar) and a position angle ${PA}=107^\circ$ (east of north) in the plane of the sky with respect to the original source of images A, B, and C. This simple model with just two point-like sources reproduces the ALMA observations well (Fig.\,\ref{fig:finalmodels}a).

As a next step, we modeled the second source as an extended one by using a S\'ersic intensity distribution profile, and played with GravLens using the VLA-15\,GHz image of \cite{sub90} (see Fig.\,\ref{fig:finalmodels}c) as a visual guide. After several trials varying its positions, size, and S\'ersic index, we found that a remarkable match between the VLA and GravLens model images (Fig.\,\ref{fig:finalmodels}b) could be obtained simply by adopting a 1D S\'ersic profile with S\'ersic index $n=1$ (i.e., an exponential intensity distribution profile), centered at the same position as the second point-like source above, and with a half-light radius of 55~mas, equal to the separation between the two point-like sources of the ALMA model. Despite its elegant simplicity and minimum set of parameters, our model successfully reproduces all the features of \PKS1830: the three compact images, the asymmetrical extensions eventually forming the Einstein ring, and the bridge-like feature between images B and C seen in the VLA image.

\section{Discussion and perspectives}

Our simple model points toward a core--jet morphology of the quasar, with a bright and compact "core" producing the three point-like images A, B, and C, and a fainter "jet" component, more extended at lower frequency, producing the extensions, the Einstein ring, and the bridge-like feature between images B and C (hereafter BC). At millimeter wavelengths, the core-jet separation ($\sim 55$~ mas) as well as the core and jet sizes ($\ll 30$~mas) are well constrained by the high angular resolution of our ALMA observations. Also at VLA frequencies, the core-jet separation cannot be much more than $\sim 50$~mas: otherwise, the extension features would extend farther out from their respective main images A and B, and even become detached if we keep a size no larger than $\sim 50$~mas for the jet. An apparent detached jet could in fact be the signature of stationary bright spots in the jet, for instance due to recollimation shocks (e.g., \citealt{bro09}).

Image C is the third image of the core, and does not correspond to the jet. This is clearly set by the ALMA observations: at millimeter wavelengths, we do not expect to detect emission from the jet because of its steep spectral index and the strong de-magnification at the location of the third image. The bridge-like feature BC, on the other hand, is due to the extended jet. Indeed, according to our lensing model, a very thin and extended ($> 0.15$~mas) jet source component toward $PA = 107^{\circ}$ would form a complete ring-like feature in the image plane, roughly centered at the middle of segment BC and with a diameter equal to the separation BC. Hence, we hypothesize that the bridge-like feature BC corresponds to the northwestern part of this ring, which is brighter as it is more favorably magnified by the lens geometry.

From our model, we derive magnifications close to three for images A and B (with opposite parity) and a de-magnification by a factor $\sim 50$ for image C (with same parity as image A). However, it should be kept in mind that the observed flux density ratios, A/B and A/C, can be modulated by the intrinsic time variability of the quasar and the time delays by up to several tens of percent. The flux density ratio A/B has indeed been seen to vary between one and two (e.g., \citealt{mul08, mar13}). Furthermore, a study of the time variations of all three images A, B, and C would bring additional evidence that C is the third image of the core.

The detection of the third image of \PKS1830, allowing us to refine the lens model of the system, opens up new perspectives:
\begin{itemize}
\item  Our model predicts time delays of $t_{\rm AB} \sim 26-29$~days and $t_{\rm AC} \sim 31-34$~days, depending on \H0 (see Table~\ref{tab:lensing-model_pointlike}). With a robust measurement of the time delay $t_{\rm AB}$, \PKS1830\ could be used in a sample of strong gravitationally lensed systems to constrain \H0\ (e.g., \citealt{won20}).
\item It would be interesting (although certainly challenging) to search for absorption toward the weak extensions and image C to obtain further insight into the content, distribution, and kinematics of the gas in the lens galaxy (e.g., \citealt{mul14}). This would help to constrain the properties of the galaxy, such as its inclination and disk rotation, which have so far only been investigated with unresolved H\,I absorption data (\citealt{koo05}). 
\item The time variability and polarization properties of the continuum emission have been used to study the different components of the quasar (e.g., \citealt{mar13, mar19, mar20}). A high-angular-resolution monitoring of \PKS1830\ (e.g., with ALMA or mm-VLBI), resolving the third image and the jet extensions, would allow us to further characterize the structure of the jet.
\end{itemize}

\begin{acknowledgement}
We thank the referee for comments and suggestions that improved the clarity of the manuscript. We would like to thank C. Keeton for his kind help with questions related to GravLens. This paper makes use of the following ALMA data: ADS/JAO.ALMA\#2018.1.00051.S. ALMA is a partnership of ESO (representing its member states), NSF (USA) and NINS (Japan), together with NRC (Canada) and NSC and ASIAA (Taiwan) and KASI (Republic of Korea), in cooperation with the Republic of Chile. The Joint ALMA Observatory is operated by ESO, AUI/NRAO and NAOJ. This research has made use of NASA's Astrophysics Data System. IMV thanks the Generalitat Valenciana for funding, in the frame of the GenT Project CIDEGENT/2018/021.
\end{acknowledgement}

\begin{appendix}

\section{Additional information} \label{app:extrainfo}

\begin{table*}[ht!]
\caption{Summary of the observations.}
\label{tab:obsdata}
\begin{center} \begin{tabular}{cccccccccc}
\hline \hline
Tuning & Frequency & Date & $N_{\rm ant}$ & $B_{\rm min}$ & $B_{\rm max}$  & PWV & $t_{\rm on}$ & Beam / P.A. & LRAS \\
       & (GHz)     & &             & (m)         & (km)         & (mm) & (min)      & (mas $\times$ mas / deg) & ($\arcsec$) \\
\hline
B4 & 154.89 & 2019 Jul 10 & 45 & 138 & 13.9 & 2.1 &  7:36 & $46.9 \times 37.7$ / 61.5 & 1.4\\ 
B5 & 181.40 & 2019 Jul 28 & 43 &  92 &  8.5 & 0.6 & 17:14 & $56.4 \times 51.6$ / 87.0 & 1.8 \\ 
B6 & 230.89 & 2019 Jul 11 & 44 & 111 & 12.6 & 1.2 & 24:52 & $37.5 \times 23.2$ / 70.0 & 1.2 \\ 
B7 & 283.38 & 2019 Jul 28 & 42 &  92 &  8.5 & 0.6 & 18:49 & $34.3 \times 32.9$ / 42.3 & 1.2 \\ 
\hline
\end{tabular} \end{center}
\tablefoot{N$_{\rm ant}$: number of antennas in the array; B$_{\rm min}$ and B$_{\rm max}$: minimum and maximum baseline, respectively; PWV: precipitable water vapor; t$_{\rm on}$: on-source observing time; Beam / P.A.: synthesized beam size and position angle; LRAS: largest recoverable angular scale, calculated as $c/(2\nu B_{\rm min})$.}
\end{table*}

\begin{table}[ht]
\caption{Flux density ratios measured during the different ALMA observations in July 2019.}
\label{tab:FluxRatios}
\begin{center} \begin{tabular}{ccccc}
\hline
Date & Band & $f_{\rm A}$ (Jy) & A/B & A/C \\
\hline
10/07/2019 & B4 & 2.85 & $0.986 \pm 0.002$ & $161 \pm 15$ \\
11/07/2019 & B6 & 2.13 & $0.965 \pm 0.002$ & $128 \pm 7$ \\
28/07/2019 & B5 & 2.48 & $1.097 \pm 0.002$ & $152 _{-11}^{+14} $ \\
28/07/2019 & B7 & 1.89 & $1.109 \pm 0.002$ & $161 \pm 14$ \\
\hline
\end{tabular} \end{center} \end{table}

\begin{table*}[ht]
\caption{Astrometry of sources in the field of  \PKS1830.} \label{tab:astrometry-literature}
\begin{center} \begin{tabular}{lccc}
\hline
Reference & Instrument & \multicolumn{2}{c}{Position relative to image A ($-$R.A., $-$Dec. offsets, mas) $^{(a)}$} \\
          &            & Image B & Others \\
\hline
\hline
\smallskip \cite{leh00} & HST (K,H,I) & (653, 721) $\pm 1$ & lens G: (501, 445) $\pm 80$  \\
\cite{cou02} & HST \& Gemini & (654, 725) $\pm 2$ & lens G: (519, 511) $\pm 80$ \\
\smallskip             & (V,I,K) &                       & SP: (285, 722) $\pm 40$ \\
\smallskip \cite{win02} & HST (V,I) &                    & lens G: (328, 486) $\pm 4$ \\

\smallskip \cite{mey05} & VLT (J,H,Ks) & (649, 724) $\pm 1$ & lens G: (498, 456) $\pm 4$ \\
             & &                    & star P: (333, 504) $\pm 2$ \\
\smallskip \cite{jin03} & VLBA (43 GHz) & (642.092, 728.086) $\pm 0.2$(systematics) $^{(b)}$ & \\             
This work & ALMA & (641.946 $\pm 0.042$, 728.082 $\pm 0.033)$ & C: (500.1 $\pm 2.1$, 459.6 $\pm 1.6$) \\
\hline
\end{tabular}
\end{center}
\tablefoot{$a)$ Offsets in R.A and Dec. are given in absolute value with respect to image A. $b)$ \cite{jin03} observed time changes in the separation between images A and B of up to 0.2~mas within eight months, most likely produced by intrinsic changes in the brightness distribution of the background quasar which is enhanced by the lens magnification. The given relative position corresponds to the nominal reference at the first epoch of their observations.}
\end{table*}

\begin{figure}[h] \begin{center}
\includegraphics[width=8.8cm]{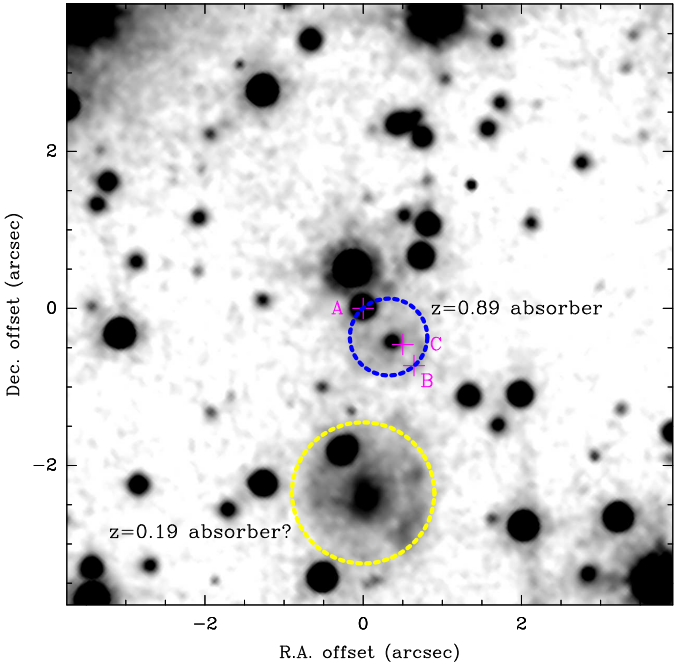}
\caption{I-band image of the surrounding of \PKS1830 obtained from the Hubble Space Telescope Archive (WFPC2, F814W; see also publications by \citealt{win02,cou02}). The three images, A, B, and C, are marked with magenta crosses. The dashed yellow circle shows the location of the candidate $z=0.19$ H\,I/OH absorber, about $2.5\arcsec$ south of image~A.}
\label{fig:HST}
\end{center} \end{figure}

\section{Fitting the lens model} \label{app:lensmodel}

\subsection{The $z=0.89$ lens} \label{app:lens}

The lensed images of a source are the solutions to the lens equation, given the configuration of source and lens positions, and lens potential:
\begin{equation} {\bf u} = {\bf x} - \alpha({\bf x}), \end{equation}
\noindent where {\bf u} and {\bf x} are the source and images positions (i.e., 2D vectors on the plane of the sky), respectively, and $\alpha$ is the deflection angle due to the 2D gravitational potential $\phi$ of the lens: 
\begin{equation}
\alpha({\bf x}) = \nabla \phi({\bf x}).
\end{equation}
The potential is linked to the dimensionless convergence $\kappa$:
\begin{equation}
\nabla^2 \phi({\bf x}) = 2 \kappa({\bf x}),
\end{equation}
which is defined as the surface density of the lens $\Sigma({\bf x})$ in units of the critical surface density $\Sigma_{cr}$ of the system:
\begin{equation}
\Sigma_{cr} = \frac{c^2}{4\pi G} \frac{D_{os}}{D_{ol}D_{ls}}. 
\end{equation}
\noindent There, $D_{os}$ is the angular distance between the observer and the source, $D_{ol}$, the angular distance between the observer and the lens, and $D_{ls}$, the angular distance between the lens and the source. In the case of the \PKS1830 system (with a flat Universe defined by $\Omega_m=0.3$ and $\Omega_{\Lambda}=0.7$), we get $\Sigma_{cr} = 2.24 \times 10^{9}$~M$_{\odot}$\,kpc$^{-2}$.

For the lens mass model, we adopt a softened isothermal ellipsoidal (see, e.g., \citealt{kor94}):
\begin{equation} \label{eq:ellipsoidal}
\kappa(x',y') = \frac{\mathcal{M}}{2} \left [ s^2 + x'^2 + \frac{y'^2}{1-e^2} \right ]^{-\frac{1}{2}}
,\end{equation}
\noindent where $\mathcal{M}$ is a mass scale, $s$ is the softening length, and $e$ is the ellipticity, with the coordinate axes $(x',y')$ centered on the lens with a position angle $\theta$.

\begin{figure*}[h] \begin{center}
\includegraphics[width=\textwidth]{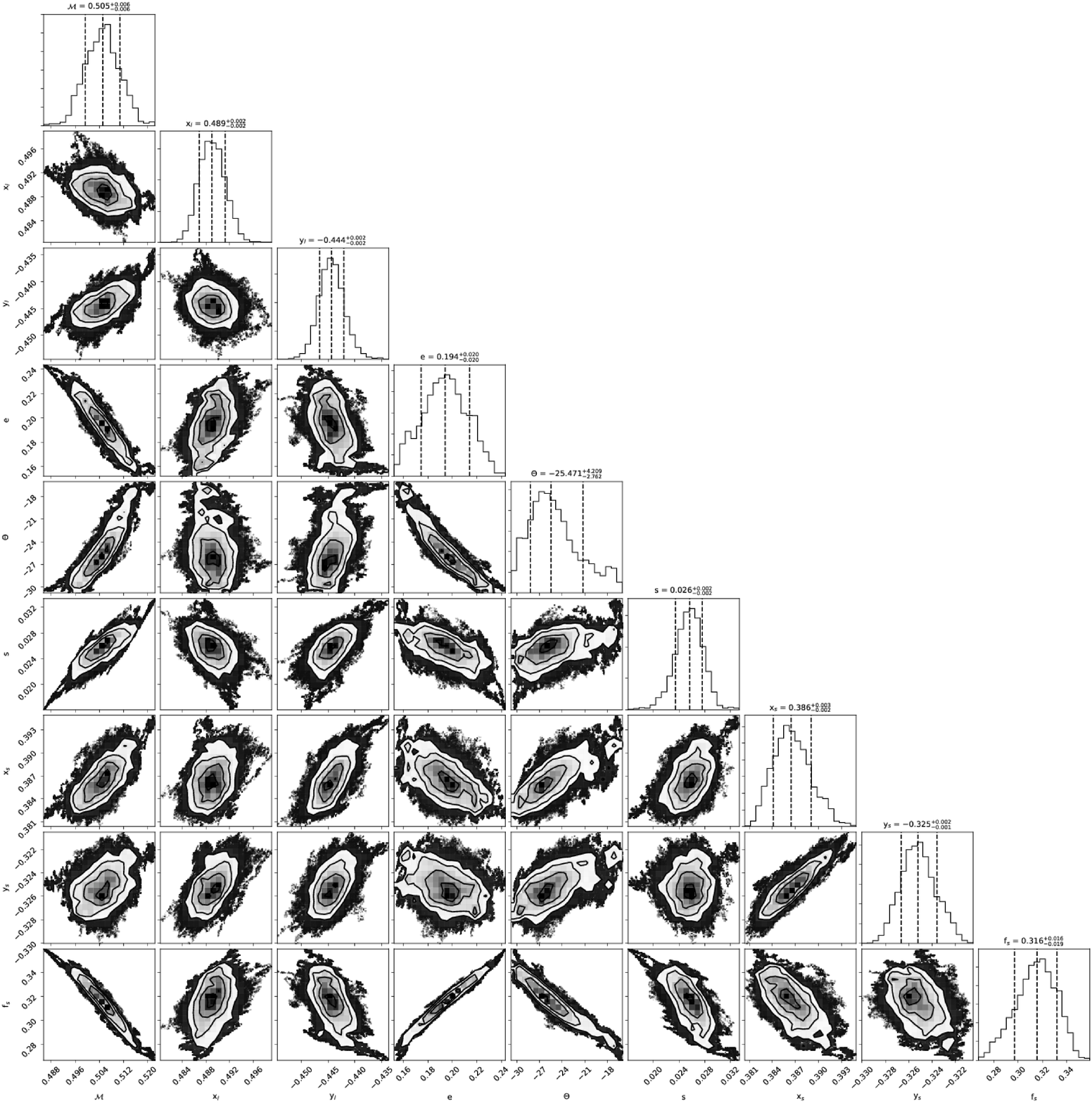}
\caption{Plots showing the parameter degeneracies from the GravLens Monte-Carlo fitting with a single point-like source. $x_s$ and $y_s$ are the position of the source (in arcsec), with respect to image A, and $f_s$ its flux normalized to that of image A. $x_l$ and $y_l$ are the positions of the center of the adopted softened isothermal ellipsoidal model (in arcsec), $e$  is the ellipticity, and $\theta$ the position angle (in degrees). $\mathcal{M}$ is the mass scale and $s$ the softening length (see Eq.\,\ref{eq:ellipsoidal}), both in arcsec.}
\label{fig:mcmc}
\end{center} \end{figure*}

Because the redshifts of the lens and source are observationally set, there are six free parameters left for the lens ($x_l$, $y_l$, $\mathcal{M}$, $s$, $e$, and $\theta$), which can be solved from the relative positions of the three lensed images. The best-fit values of those parameters are given in Table~\ref{tab:lensing-model_pointlike}. Corner plots are shown in Fig.\,\ref{fig:mcmc}.

From the monopole deflection for our lens model, we can calculate the mass enclosed within a given radius, which we find to be $\sim 4 \times 10^{11}$\,M$_\odot$ within a region of $1\arcsec$ in diameter (i.e., $\sim 7.4$\,kpc) centered on the lens. Within a radius of 40\,kpc, we estimate a total enclosed mass of roughly $9 \times 10^{11}$\,M$_\odot$.
As expected, the position of the lens center is very close ($< 20$~mas) to that of the third image.

\begin{table*}[ht]
\caption{Input parameters and results of the GravLens model fitting of the lens.}
\label{tab:lensing-model_pointlike}
\begin{center} \begin{tabular}{llccl}
\hline \hline
\multicolumn{5}{c}{\em Input parameters}\\
\hline
Cosmology & $\Omega_m$, $\Omega_\Lambda$ & 0.3, 0.7 & & \\
          & \H0\ (\kms\,Mpc$^{-1}$) & 67.4$\pm$0.5; $73.3_{-1.8}^{+1.7}$ & & (1) \\
\hline
Source & Redshift $z_S$ & 2.507 & & (2) \\
\smallskip        & Intensity profile & point-like & & \\
Lens & Redshift $z_L$ & 0.88582 & & (3) \\
     & Initial guess position & -500.1, -459.6 & & (4) \\
\smallskip    & Mass distribution & softened isothermal ellipsoidal & & \\
Lensed images & Position of image A & $(\Delta {\rm RA}, \Delta {\rm Dec})\equiv (0,0)$ & & (5) \\ 
              & Position of image B (mas) & $-641.946 \pm 0.042, -728.082 \pm 0.033$ & & (6)  \\
              & Position of image C (mas) & $-500.1 \pm 2.1, -459.6 \pm 1.6$ & & (6) \\
              & Flux ratios (A:B:C) & $1:0.93\pm 0.09:0.0071\pm0.0009 $ & & (7)  \\
\hline
\multicolumn{5}{c}{\em Output parameters} \\
& & Best fit & Guess range & \\
\hline
Source     & Position $x_s,y_s$ (mas) & $-386 \pm 3, -325 \pm 2$ &&  (8) \\
\smallskip & Flux $f_s$ & $0.317 \ ^{+0.016}_{-0.019}$ && (9) \\
Lens        & Position $x_l,y_l$ (mas) & $-489 \pm 2, -444 \pm 2$ & &  (8) \\
            & Mass scale (mas) & $505 \pm 6$ & [100; 1500] & (10) \\
            & Softening length $s$ (mas)& $26 \pm 2$  & [10; 200] & (10) \\
            & Ellipticity $e$ & $0.19 \pm 0.02$ & [0.0; 0.3] & \\
            & Position angle $\theta$ ($^{\circ}$) & $-25 \ ^{+4}_{-3}$ & [$-90$;+90]& \\
\hline
\multicolumn{5}{c}{\em Derived quantities} \\
\hline
Time delay  & $t_{\rm AB}$ (d) & 28.6; 26.3 & & (11) \\
\smallskip  & $t_{\rm AC}$ (d) & 33.7; 31.0 & & (11) \\
Magnification & $\mu_{\rm A}$ & 3.1 & & \\
                       & $\mu_{\rm B}$ & $-2.9$ & &  \\
                       & $\mu_{\rm C}$ & $0.022$ & & \\ 
\hline
\end{tabular}\end{center}
\tablefoot{ (1) $H_0$ measurements from \cite{Planck18} and \cite{won20}, respectively.
(2) \cite{lid99}.
(3) \cite{wik96}.
(4) Set as image C (Table~\ref{tab:info-images}), as both are expected to be very close in the plane of the sky.
(5) Taken as position reference.
(6) Positions from this work (Table~\ref{tab:info-images}) given with respect to image A.
(7) Flux ratios from this work (Table~\ref{tab:info-images}).
(8) Positions relative to image A. 
(9) Flux of the source normalized to that of image A (i.e., inverse of the magnification, $f_s = 1/\mu_{\rm A}$).
(10) As defined in Eq.\,\ref{eq:ellipsoidal}.
(11) Time delays corresponding to $H_0$ values of 67.4~\kms\,Mpc$^{-1}$ and 73.3~\kms\,Mpc$^{-1}$, respectively.
}
\end{table*}

\subsection{The second absorber at $z=0.19$} \label{app:2ndabsorber}

There is a second absorber at a redshift $z=0.19$ in the line of sight of \PKS1830, detected in H\,I and OH absorption, first by \cite{lov96} and re-observed more recently by \cite{all17}. The closest galaxy in the immediate surroundings of \PKS1830\ is located about $2.5\arcsec$ south of image A, as can be seen in the HST image shown in Fig.\,\ref{fig:HST} (see also \citealt{leh00,win02,cou02,mey05} for analysis and discussion of optical images of \PKS1830). To our knowledge there is no spectroscopic confirmation that this galaxy is indeed the second absorber but there are no other obvious candidates. The galaxy is brighter than the $z=0.89$ galaxy and has a larger angular size, and its photometry is consistent with that of a relatively nearby spiral (\citealt{win02}). 

We can calculate the convergence $\kappa'$ due to an object at the location of this second galaxy for a range of redshift and mass. Assuming a singular isothermal sphere model for the new lens galaxy with 1D stellar dispersion $\sigma_v$, we have
\begin{equation}
\kappa'(\theta) = \frac{2\pi \sigma_v^2}{c^2}\frac{D_{ls}}{D_{os}}\frac{1}{\theta},
\end{equation}
\noindent where $\theta$ is the angular distance between the new lens and a given location. We find that its lensing effect is negligible at the location of both images A and B (slightly larger at image B because it is closer in the sky to the second galaxy), with $\kappa' < 1$ (and decreasing with increasing redshift of the object) as long as its stellar dispersion $\sigma_v < 350$~\kms. For $z = 0.2$ and $\sigma_v = 200$~\kms, $\kappa' = 0.3$ at image B. Therefore, only a nearby and very massive galaxy could introduce some weak lensing effect.

\subsection{The source} \label{app:source}

The first lens model was derived considering a single point-like source $S_1$ at the origin of the three compact lensed images A, B, and C. To account for the extensions, one needs to add source components.

To reproduce the weak and compact extensions seen on the ALMA observations, we simply added one more point-like component $S_2$ (in practice a S\'ersic intensity profile with a very small half-light radius of 1~mas), and found that it should be set at a position 55~mas with a position angle ($PA$) of 107$^{\circ}$ with respect to the first source $S_1$.

To further reproduce the partial Einstein ring as mapped with the VLA (e.g., \citealt{sub90}), we need to move the second source into an extended component $S_2'$, for example using a S\'ersic intensity profile. A simple but remarkably good solution was found by setting $S_2'$ at the same position as $S_2$, using a 1D S\'ersic profile with a half-light radius of 55~mas (i.e., the distance between $S_1$ and $S_2$), and a S\'ersic index $n=1$, i.e., corresponding to an exponential profile (Table~\ref{tab:ext_sou_models}). To account for the sharp cut in emission to the northwest of image A and southeast of image B in the VLA images, one should suppress emission opposite to the jet. However, the match between our model and the observations is already more than satisfying for the purpose of this study, especially given the simplicity (small number of parameters) of our model. This is also why we refrained from exploring further ellipsoidal (2D) S\'ersic intensity profiles for the extended source.

The parameters of the ALMA and VLA source models (Fig.\,\ref{fig:finalmodels}) are given in Table~\ref{tab:ext_sou_models}.

\begin{table*}[ht]
\caption{Parameters of the S\'ersic intensity profiles used for the ALMA and VLA source models (Fig.\,\ref{fig:finalmodels}a and b, respectively).}
\label{tab:ext_sou_models}
\begin{center} \begin{tabular}{llcccc}
\hline \hline
           &    & (X,Y) $^{(a)}$ & $r_0$ $^{(b)}$ & $I_0$ $^{(c)}$ & $n$ $^{(d)}$ \\
           &    & (mas) & (mas) &       & \\
\hline
ALMA model & $S_1$ & $(-386,-325)$ & 1 & 1 & 1 \\
           & $S_2$ & $(-333,-341)$ & 1 & 0.005 & 1 \\

VLA model  & $S_1$ & $(-386,-325)$ & 1 & 1 & 1 \\
          & $S_2'$ & $(-333,-341)$ & 55 & 0.5 & 1 \\
\hline
\end{tabular}\end{center}
\tablefoot{$a)$ Positions are given with respect to image A;
$b)$ Half-light radius;
$c)$ Intensity;
$d)$ S\'ersic index.
}
\end{table*}

\end{appendix}
\end{document}